\documentclass{article}
\usepackage{ijcai23}

\usepackage{times}
\usepackage{soul}
\usepackage{url}
\usepackage[hidelinks]{hyperref}
\usepackage[utf8]{inputenc}
\usepackage[small]{caption}
\usepackage{graphicx}
\usepackage{amsmath}
\usepackage{amsthm}
\usepackage{booktabs}
\usepackage{algorithm}
\usepackage{algorithmic}
\usepackage[switch]{lineno}
\usepackage{amsfonts}
\usepackage[table,xcdraw]{xcolor}
\usepackage{multirow}
\usepackage{tablefootnote}

\title{Signed Graph Representation Learning: A Survey}

\author{
Zeyu Zhang$^1$
\and
Peiyao Zhao$^2$\and
Xin Li$^2$\and 
Jiamou Liu$^3$\and \\
Xinrui Zhang$^4$\and
Junjie Huang$^4$\And
Xiaofeng Zhu$^5$
\affiliations
$^1$Huazhong Agricultural University\\
$^2$Beijing Institute of Technology\\
$^3$The University of Auckland\\
$^4$Southwest University\\
$^5$University of Electronic Science and Technology of China
\emails
zhangzeyu@mail.hzau.edu.cn,
\{peiyaozhao, xinli\}@bit.edu.cn,
jiamou.liu@auckland.ac.nz,
zhangxinrui@email.swu.edu.cn,
junjiehuang.cs@outlook.com,
seanzhuxf@gmail.com,
}

\begin{document}

\maketitle

\begin{abstract}
With the prevalence of social media, the connectedness between people has been greatly enhanced. 
Real-world relations between users on social media are often not limited to expressing positive ties such as friendship, trust, and agreement, but they also reflect negative ties such as enmity, mistrust, and disagreement, which can be well modelled by signed graphs. 
Signed Graph Representation Learning (SGRL) is an effective approach to analyze the complex patterns in real-world signed graphs with the co-existence of positive and negative links. 
In recent years, SGRL has witnesses fruitful results. 
SGRL tries to allocate low-dimensional representations to nodes and edges which could preserve the graph structure, attribute and some collective properties, e.g., balance theory and status theory. 
To the best of knowledge, there is no survey paper about SGRL up to now. 
In this paper, we present a broad review of SGRL methods and discuss some future research directions.
\end{abstract}


\section{Introduction}

The widespread impact of online platforms, like social media, business dealings, and cryptocurrency transactions, has led to a huge increase in graph datasets. 
These datasets, with complex and connected structures, pose a big challenge for analysis. 
ver the past decade, graph machine learning methods, especially Graph Neural Networks (GNNs) \cite{kipf2016semi,hamilton2017inductive,velivckovic2017graph}, have gained increasing attention in both academic and industrial circles. These approaches have significantly advanced in various applications, such as link prediction, node classification, and graph classification. 

\begin{figure}
    \centering
    \includegraphics[width=0.5\textwidth]{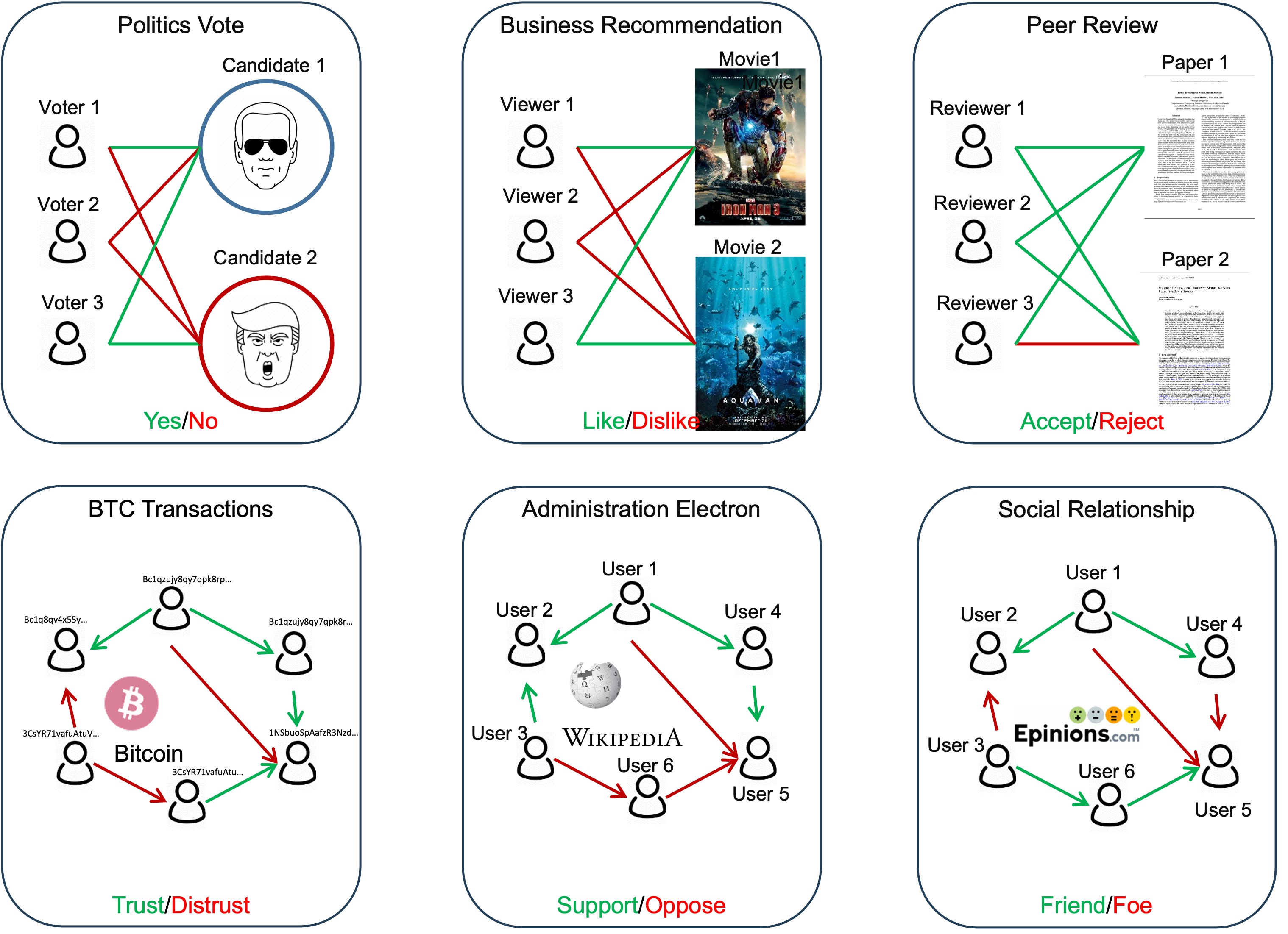}
    \caption{An illustration of signed graph in real world.}
    \label{fig:fig-1}
    \vspace{-1em}
\end{figure}

Despite significant progress in GNNs, most GNN methods are designed for unsigned graphs (consisting of only positive edges). However, real-world edge connections among nodes frequently extend beyond conveying positive relationships like friend, accept, trust, and support. 
They may also encompass negative associations, such as foe, rejection, distrust, and opposition. 
In Figure \ref{fig:fig-1}, we find that signed graphs have numerous applications in our daily lives, spanning various domains and contexts.
This setup can be thought of as a graph with both positive and negative connections. 
However, having negative links messes up the usual way information is passed, so we need new signed graph representation learning (SGRL) models, like Signed Graph Neural Networks (SGNN), to handle this mix of positive and negative connections.

For unsigned graph representation learning, there exist several surveys in recent years \cite{cui2018survey,cai2018comprehensive,wu2020comprehensive,zhang2020deep,zhou2020graph}. 
\cite{cai2018comprehensive} provides a review of graph embedding methods and classifies the current methods into matrix factorization, deep learning, matrix reconstruction, graph kernel, generative models, and so on.
\cite{cui2018survey} focuses on categorizing and reviewing current network embedding methods including linear models and deep models. 
\cite{wu2020comprehensive} only focuses on deep models of graph representation learning and proposes a new taxonomy to classify GNN models into four categories, i.e., recurrent GNNs, convolutional GNNs, graph autoencoder, and spatial-temporal GNNs. 
Similar to \cite{wu2020comprehensive}, \cite{zhang2020deep} also presents a comprehensive review of different types of deep learning methods on graphs. 
\cite{zhou2020graph} provides an overview of GNN variants from different angles, i.e., graph type, training methods, and propagation step. 
In addition to these surveys, there exist several related surveys about the extended application of graph neural networks \cite{lamb2020graph,wu2020graph}.
For signed graph mining, one of the important surveys is \cite{tang2016survey}. 
This survey gives an overall review of signed graph properties and analysis tasks but does not consider graph representation learning methods as this direction only begins to receive attention in recent years. 
Another signed graph-related survey is \cite{zheng2015social}. This survey provides a comprehensive overview of significant research regarding social balance, one of the important signed graph collective properties, including fundamental measures, detecting algorithms, and so on.


In summary, to the best of our knowledge, no surveys are specifically written for the SGRL methods. 
Thus, in this paper, we provide a comprehensive and unique survey about SGRL to bridge the gap. 
We believe this survey can not only help researchers gain a deep understanding of the current progress of SGRL but also provide rich resources and potential future directions in this field.

\section{Challenges}
Despite significant progress in graph representation learning, research on SGRL remains crucial for the following challenges:


\textbf{How to handle negative edges}. Most Graph Neural Networks (GNNs) for unsigned networks use homophily, assuming connected users share similarities. However, this concept may not directly apply to signed networks, where negative edges represent opposing semantics.


\textbf{Limitation of balance theory}. Many SGRL methods rely on balance theory, rooted in psychology, which simplistically divides nodes into two conflicting groups. However, this falls short of capturing the complexity of real signed graphs. It is essential to explore extensions of current SGNN methods to handle cases with more than two groups (k-groups).

\textbf{Representation limitation of SGNN models}. Unbalanced triangles, common in real-world signed graph datasets, challenge current SGNN methods in accurate representation learning \cite{zhang2023rsgnn}. Increased unbalanced triangles due to factors like random noise and adversary attacks impact predictive performance, highlighting the need to explore mitigation strategies.


\textbf{Diverse learning tasks}. Downstream tasks for signed graph representation, like link sign prediction and node classification, differ from unsigned graphs. Node classification is particularly challenging in signed graph datasets without node labels and attributes, relying solely on the network.


\section{Preliminaries and Problem Formulation}

This section will introduce some preliminaries to lay the foundations for further discussion, including the preliminaries for signed networks, popular signed GNN models, and basic signed social theory.

\subsection{Signed Social Graphs}
In this subsection, we introduce notations and give the necessary backgrounds on the signed graph.

\textbf{Notations}. We denote a signed graph as $\mathcal{G}=\{\mathcal{V}, \mathcal{E}, s\}$, where $\mathcal{V}$ is the node sets in a signed graph $\mathcal{G}$, and $\mathcal{E} \subset \mathcal{V}\times \mathcal{V}$ represents the sets of edges in graph $\mathcal{G}$ with sign $s$. 
$\mathcal{E}^+$ and $\mathcal{E}^-$ denote the sets of positive and negative links, respectively. It means that there is at most one signed edge between any two nodes.
$\mathcal{E}$ contains $\mathcal{E}^+$ and $\mathcal{E}^-$. Similarly, we can divide the neighbors of node $v_i$ into positive neighbors $\mathcal{N}_i^+$ and negative neighbors $\mathcal{N}_i^-$.
Also, $\mathcal{G}$ can be denoted as the adjacency matrix $A$, where $A_{ij} = 1$ means there exists a positive link between $v_i$ and $v_j$, $A_{ij}=-1$ denotes a negative link between $u_i$ and $u_j$, and $A_{ij}=0$ means there is no link between $u_i$ and $u_j$.
Usually, $s\in \{1, -1\}$ is the sign of the edge\footnote{The signed network can be further associated with weights. \cite{kumar2016edge} } and $\mathcal{E}^+ \bigcap \mathcal{E}^- = \emptyset$; 

\textbf{Problem formulation}. Given a signed graph $\mathcal{G}=\{\mathcal{V}, \mathcal{E}, s\}$, SGRL aims to encode the nodes $\mathcal{V}$ into low-dimensional vectors $Z\in \mathbb{R}^{d \times |\mathcal{V}|}$ as follows:
\begin{equation}
    f(\mathcal{G}) \rightarrow Z,
\end{equation}
where $d$ is the embedding size ($d \ll |\mathcal{V}|$), and $f$ is the encoder function which can be Network Embedding or GNNs. 
The low-dimensional vectors $Z$ should be useful in downstream machine-learning tasks.

\begin{table}
\centering
\resizebox{\linewidth}{!}{
\begin{tabular}{llllll}
\toprule
Dataset              & \# Nodes     & \# Edges        & \%Pos  & Sign Semantics & Scenarios \\ 
\midrule
Bitcoin-Alpha    & 3,783   & 24,186    & 93.7 & Trust/Distrust  & \multirow{6}{*}{\textit{Social Relationship}}\\
Bitcoin-OTC      & 5,881   & 35,592    & 90.0 & Trust/Distrust\\
Wikirfa          & 11,259  & 178,096   & 77.9 & Support/Oppose \\
Slashdot         & 82,144  & 549,202   & 77.4 & Like/Dislike\\
Epinions         & 131,828 & 841,372   & 85.3 & Trust/Distrust\\
RedditHyperlinks & 55,863  & 858,490   & 92.7 & Positive/Negative\\
Bonanza          & 9,892   & 36,543    & 98.0 & Positive/Negative & \multirow{4}{*}{\textit{Object Opinion}} \\
U.S. House       & 1,796   & 114,378   & 54.0 & Support/Oppose \\
U.S. Senate      & 1,201   & 27,083    & 44.7 & Support/Oppose\\
PeerReview       & 486     & 1,170     & 39.7 & Accept/Reject\\ 
ML-1M            & 9,992   & 1,000,209 & 57.5 & High-rate/Low-rate & \multirow{5}{*}{\textit{Interaction Feedback}}\\
Amazon-Book      & 73,857  & 1,960,674 & 80.6 & High-rate/Low-rate \\
Yelp             & 66,418  & 1,900,308 & 68.2 & High-rate/Low-rate\\
Zhihu            & 17,437  & 554,150   & 29.3 & Interact/Skip\\
WeChat           & 8,582   & 13,557    & 85.7 & Read/Report\\
\bottomrule
\end{tabular}
}
\caption{Common Datasets for Signed Networks.}
\label{tab:dataset}
\end{table}

\textbf{Datasets}. Beyond social relationships, antagonistic relationships also exist in complex systems like diplomacy, biology, and chemistry. Examples include cooperative and hostile relations in international affairs, excitatory and inhibitory relationships among neurons in biology \cite{tang2016survey}, and the side effects between chemical drugs \cite{zitnik2018modeling}. 
While these complex systems also feature signed links, this paper primarily focuses on signed links in social relationships, where the entities involved are individuals.
We have collected and organized datasets commonly used for modeling signed networks, as shown in Table \ref{tab:dataset}. 
From Table \ref{tab:dataset}, it can be seen that the modeling of signed networks can be categorized into three different types based on the context of the links. Thus, the scenarios of signed networks are divided into: \textit{social relationships} between people, \textit{object opinions} towards specific topics, and \textit{interaction feedback} in human-computer interaction.

\textbf{Machine learning tasks}
Common graph machine learning tasks, as detailed in \cite{wu2020comprehensive}, typically fall into three distinct categories: Node-level, Link-level, and Graph-level. 
Node-level graph machine-learning tasks usually involve applying machine learning to predict nodes' attributes. 
Link-level tasks primarily focus on predicting the relationships between edges in the network\footnote{
In signed graphs, link prediction typically includes predicting the sign of a given link (binary classification) and predicting both the existence and the sign of a link (3-class classification).}. 
In graph-level machine learning tasks, there are generally two main subjects of study: subgraph-level and whole-graph-level.
At the subgraph level, the focus is on clustering nodes within social networks, with common tasks including Node Clustering \cite{he2022sssnet} and Community Detection \cite{sun2020stable}. 
Whole-graph-level tasks mainly encompass Graph Classification \cite{liu2022graph} and Graph Regression\cite{li2019senti2pop}.
It is worth mentioning that the mainstream machine learning task of SGRL is the Link-level.

\subsection{Signed Graph Neural Network}

The architecture of current popular SGNN models, such as SGCN \cite{derr2018signed} and SNEA \cite{li2020learning} follows the spatial unsigned GNNs utilizing neural message passing among nodes to define convolutions on graphs, but adopt the following mechanism.
The representation of a node $v_i$ at a given layer $\ell$ is defined as
\[
h_i^{(\ell)} = [ h_i^{pos(\ell)}, h_i^{neg(\ell)} ],
\]
where $h_i^{pos(\ell)}$ and $ h_i^{neg(\ell)}$ respectively denote the positive and negative representation vectors of node $v_i \in \mathcal{V}$ at the $\ell$th layer, and $[,]$ denotes the concatenation operation. 
The unsigned GNN message passing is defined as 
\begin{equation}
\scriptsize
    h_i^{(l)}=\text{COMBINE}^{(\ell-1)}\left(h_i^{(\ell-1)},\text{AGGREGATE}^{(\ell-1)}(\{ h_j: v_j \in \mathcal{N}_i \})\right),
\end{equation}
where $\text{COMBINE}$ and $\text{AGGREGATE}$ are differentiable functions (e.g., Mean, Max). 
Different from GNNs, SGNNs accommodate positive and negative edges using a two-part representation, and a more involved aggregation scheme. For example, when $\ell>1$, the positive part of the representation for node $v_i$ could aggregate information from the positive-representation of its positive neighbors and the negative-representation of its negative neighbors.
\begin{equation}
\label{eq:SGNN_2}
\tiny	
\begin{aligned}
    h_i^{pos(\ell)} = & \text{COMBINE}^{(\ell)} \Biggl(h_i^{pos(\ell-1)}, \\
    & \text{AGGREGATE}^{(\ell)} ( \left\{h_j^{pos(\ell-1)}: v_j \in \mathcal{N}_i^+ \right\},\left\{h_j^{neg(\ell-1)}: v_j \in \mathcal{N}_i^- \right\} ) \Biggl) \\
    h_i^{neg(\ell)} = & \text{COMBINE}^{(\ell)} \Biggl(h_i^{neg(\ell-1)}, \\
     & \text{AGGREGATE}^{(\ell)} ( \left\{h_j^{neg(\ell-1)}: v_j \in \mathcal{N}_i^+ \right\}, \left\{h_j^{pos(\ell-1)}: v_j \in \mathcal{N}_i^- \right\} ) \Biggl), \\
\end{aligned}
\end{equation}
After $\ell$ SGNN layers, we have the final representation $Z_i=\text{CONCATENATE}(h_i^{\ell}, h_i^{\ell})$. 
In addition to $\text{CONCATENATE}$, we can also use $\text{MLP}$ or $\text{ATTENTION}$ mechanisms to fuse positive and negative representations.

%

\subsection{Social Theory}
In the realm of traditional social network theory, a considerable amount of research has focused on the concepts of Homophily and Heterophily. 
However, when negative links are introduced into social networks, many assumptions no longer hold.
For example, heterophily assumptions also suggests that connected nodes are prone to have different properties or labels. 
However, it significantly differs from \textbf{Balance theory} in its inability to capture certain concepts like ``an enemy of an enemy is a friend". 
In this section, we introduce two important sociological theories that are more prevalent in signed graphs: \textbf{Balance theory} and \textbf{Status theory}.

\begin{figure}
    \centering
    \includegraphics[width=\linewidth]{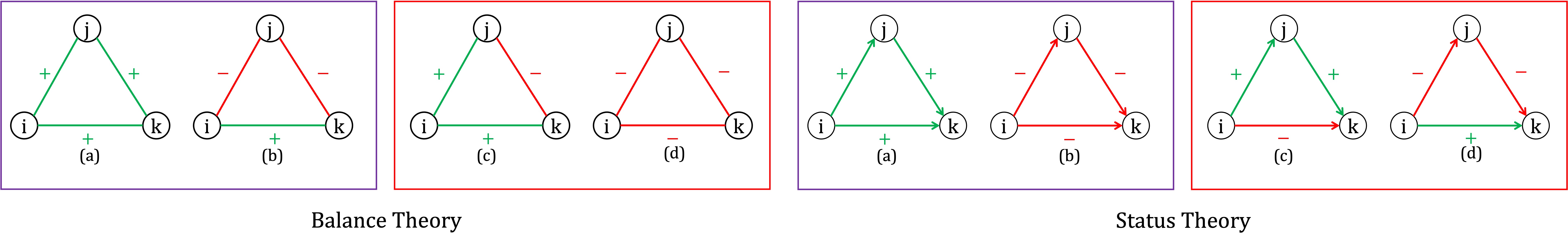}
    \caption{Four isomorphism types of triads. Blue and red lines represent positive and negative edges, resp.}
    \label{fig:triad}
\end{figure}

\textbf{Balance theory} \cite{heider1946attitudes}, a concept in social psychology, was developed by Fritz Heider in the 1940s. 
The theory focuses on the idea that individuals seek cognitive consistency and balance in their attitudes and relationships within social networks. 
According to balance theory, people strive to maintain a psychological balance in their interactions. 
In the context of triadic relationships, where three entities are connected by positive and negative ties, balance theory suggests that individuals tend to prefer balanced configurations. 
For example, if $v_i$ has a positive relationship with $v_j$ and a negative relationship with person $v_k$ there is an imbalance in the triad. 
To restore balance, $v_i$ might either develop a positive relationship with $v_k$ or a negative relationship with person $v_j$.
Figure~\ref{fig:triad} shows all four isomorphism types of triads. 

\textbf{Status theory} \cite{leskovec2010signed} is relevant for directed networks compared to balance theory is naturally defined for undirected networks. 
Social status can be represented in a variety of ways, such as the rankings of nodes in social networks, and it represents the prestige of nodes. 
In its most basic form, \textbf{Status theory} suggests that $v_i$ has a higher status than $v_j$ if there is a positive link from $v_j$ to $v_i$ or a negative link from $v_i$ to $v_j$.
In Figure~\ref{fig:triad}, we can see the first two satisfy \textbf{Status theory}, while the latter two do not.
This status relationship is transitive.

When comparing the two theories, \textbf{Balance theory} can be understood as simulating individual preferences, like liking or disliking, and primarily applies to undirected signed graph. 
On the other hand, \textbf{Status theory} is based on the assessment of individuals' social status and is direction-dependent.
Both theories can be modeled within signed directed triads (see Figure \ref{fig:triad}). 
\cite{chen2018bridge} propose to use triplets ($i$, $j$, $k$) to comparatively measure and analyze the two theories. 
They found that in the Slashdot and Epinions datasets, approximately 75\% of the triangles satisfy both sociological theories simultaneously, and only about 1\% of triangles satisfy neither.

\begin{table*}[]
\centering
\caption{A summary of SGRL methods. ``Information Source" denotes the information used during the encoding process. ``Encoder type" refers to the foundational types on which the encoder design is based. ``Loss Type" denotes the loss function of the Optimization goal. ``Social Theory" refers to the collective properties of signed graph datasets, i.e., balance theory, status theory. ``Task" indicate the learning task of each method.}
\footnotesize
\resizebox{\textwidth}{!}{
\begin{tabular}{llllll}
\hline
\multicolumn{1}{c|}{\textbf{Method}} & \multicolumn{1}{c|}{\textbf{Information Type}}  & \multicolumn{1}{c|}{\textbf{Encoder Type}} & \multicolumn{1}{c|}{\textbf{Loss Type}}          & \multicolumn{1}{c|}{\textbf{Social Theory}} & \multicolumn{1}{c}{\textbf{Task}}    \\ \hline
\multicolumn{6}{c}{\cellcolor[HTML]{EFEFEF}Network Embedding}                                                                                                                                                                                                               \\ \hline
\multicolumn{1}{l|}{SNE \cite{yuan2017sne}}             & \multicolumn{1}{l|}{Network only}               & \multicolumn{1}{l|}{Log-Bilinear Model}    & \multicolumn{1}{l|}{MLE}                         & \multicolumn{1}{l|}{Balance Theory}         & Node Classfication,  Link Prediction \\
\multicolumn{1}{l|}{SIGNet \cite{islam2018signet}}          & \multicolumn{1}{l|}{Network only}               & \multicolumn{1}{l|}{word2vec}              & \multicolumn{1}{l|}{MLE}                         & \multicolumn{1}{l|}{Balance Theory}         & Link Sign Prediction                 \\
\multicolumn{1}{l|}{SIDE \cite{kim2018side}}            & \multicolumn{1}{l|}{Network only}               & \multicolumn{1}{l|}{Skip-Gram}             & \multicolumn{1}{l|}{MLE}                         & \multicolumn{1}{l|}{Balance Theory}         & Link Sign Prediction                 \\
\multicolumn{1}{l|}{ROSE \cite{javari2020rose}}            & \multicolumn{1}{l|}{Network only}               & \multicolumn{1}{l|}{Node2vec}              & \multicolumn{1}{l|}{MLE}                         & \multicolumn{1}{l|}{N/A}                    & Link Sign Prediction                 \\
\multicolumn{1}{l|}{SPONGE \cite{cucuringu2019sponge}}          & \multicolumn{1}{l|}{Network only}               & \multicolumn{1}{l|}{Spectral Method}       & \multicolumn{1}{c|}{Four-player minimax game}    & \multicolumn{1}{l|}{Balance theory}         & Node Classification                  \\ \hline
\multicolumn{6}{c}{\cellcolor[HTML]{EFEFEF}Deep Model}                                                                                                                                                                                                                      \\ \hline
\multicolumn{1}{l|}{SiNE \cite{wang2017signed}}            & \multicolumn{1}{l|}{Network, node feature}      & \multicolumn{1}{l|}{Neural Nework}         & \multicolumn{1}{l|}{Reconstruction}              & \multicolumn{1}{l|}{Balance theory}         & Link Sign Prediction                 \\
\multicolumn{1}{l|}{SSNE \cite{lu2019ssne}}            & \multicolumn{1}{l|}{Network only}               & \multicolumn{1}{l|}{Neural Network}        & \multicolumn{1}{l|}{Energy-based ranking}        & \multicolumn{1}{l|}{Status theory}          & Link Sign Prediction                 \\
\multicolumn{1}{l|}{BESIDE \cite{chen2018bridge}}          & \multicolumn{1}{l|}{Network Only}               & \multicolumn{1}{l|}{Neural Network}        & \multicolumn{1}{l|}{Cross Entropy}               & \multicolumn{1}{l|}{Balance/Status}         & Link Sign Prediction                 \\
\multicolumn{1}{l|}{SGCN \cite{derr2018signed}}            & \multicolumn{1}{l|}{Network, Node feature}      & \multicolumn{1}{l|}{GCN}                   & \multicolumn{1}{l|}{Cross Entropy, Structure}    & \multicolumn{1}{l|}{Balance Theory}         & Link Sign Prediction                 \\
\multicolumn{1}{l|}{SHINE \cite{wang2018shine}}           & \multicolumn{1}{l|}{Multi-Network}              & \multicolumn{1}{l|}{Autoencoder}           & \multicolumn{1}{l|}{Reconstruction}              & \multicolumn{1}{l|}{N/A}                    & Link Sign Prediction                 \\
\multicolumn{1}{l|}{SSSNET \cite{he2022sssnet}}          & \multicolumn{1}{l|}{Network, node feature}      & \multicolumn{1}{l|}{MLP}                   & \multicolumn{1}{l|}{Reconstruction}              & \multicolumn{1}{l|}{Balance Theory}         & Node Clustering                      \\
\multicolumn{1}{l|}{SNEA \cite{li2020learning}}            & \multicolumn{1}{l|}{Network, node feature}      & \multicolumn{1}{l|}{GAT}                   & \multicolumn{1}{l|}{Cross Entropy, Structure}    & \multicolumn{1}{l|}{Balance Theory}         & Link Sign Prediction                 \\
\multicolumn{1}{l|}{GS-GNN \cite{liu2021signed}}          & \multicolumn{1}{l|}{Network, node feature}      & \multicolumn{1}{l|}{Dual GCN}              & \multicolumn{1}{l|}{Cross Entropy}               & \multicolumn{1}{l|}{N/A}                    & Link Sign Prediction                 \\
\multicolumn{1}{l|}{SiGAT \cite{huang2019signed}}           & \multicolumn{1}{l|}{Network only, node feature} & \multicolumn{1}{l|}{GAT}                   & \multicolumn{1}{l|}{Reconstruction}              & \multicolumn{1}{l|}{Balance/Status}         & Link Sign Prediction                 \\ \hline
\multicolumn{6}{c}{\cellcolor[HTML]{EFEFEF}Trustworthy Model}                                                                                                                                                                                                               \\ \hline
\multicolumn{1}{l|}{ASiNE \cite{lee2020asine}}           & \multicolumn{1}{l|}{Network Only}               & \multicolumn{1}{l|}{GAN}                   & \multicolumn{1}{l|}{Four-player minimax game}    & \multicolumn{1}{l|}{Balance Theory}         & Link Sign Prediction                 \\
\multicolumn{1}{l|}{SigGAN \cite{chakraborty2023siggan}}          & \multicolumn{1}{l|}{Network Only}               & \multicolumn{1}{l|}{GAN}                   & \multicolumn{1}{l|}{Two-player minimax game}     & \multicolumn{1}{l|}{Balance Theory}         & Link Sign Prediction                 \\
\multicolumn{1}{l|}{SGCL \cite{shu2021sgcl}}            & \multicolumn{1}{l|}{Network, Node feature}      & \multicolumn{1}{l|}{Multi GAT}             & \multicolumn{1}{l|}{Cross Entropy, Contrastive}  & \multicolumn{1}{l|}{Balance Theory}         & Link Sign Prediction                 \\
\multicolumn{1}{l|}{RSGNN \cite{zhang2023rsgnn}}           & \multicolumn{1}{l|}{Network, Node feature}      & \multicolumn{1}{l|}{GCN}                   & \multicolumn{1}{l|}{Sparsity, High Balance, etc $^a$} & \multicolumn{1}{l|}{Balance/Status}         & Link Sign Prediction                 \\
\multicolumn{1}{l|}{SGA \cite{zhang2023sga}}             & \multicolumn{1}{l|}{Network Only}               & \multicolumn{1}{l|}{GCN}                   & \multicolumn{1}{l|}{Cross Entropy}               & \multicolumn{1}{l|}{Balance Theory}         & Link Sign Prediction                 \\ \hline
\multicolumn{6}{l}{%
  \begin{minipage}{6.5cm}%
    \tiny $a.$ feature loss, reconstruction and cross entropy loss
  \end{minipage}
  }
\end{tabular}
}
\end{table*}

 \section{Categorization}\label{sec:Category}
To transform signed graphs from topology space to embedding spaces, there are different methods adopted in current models. In this section, We categorize signed signed graph representation methods into three parts, i.e., network embedding methods, deep models and trustworthy models. 



\subsection{Shallow Network Embedding}
Shallow network embedding typically refers to those methods with simple, linear non-deep learning structures. This type of method is mainly divided into two aspects: methods based on random walks (\textbf{SNE} \cite{wang2017signed}, \textbf{SIGNet} \cite{islam2018signet}, \textbf{SIDE} \cite{kim2018side}, \textbf{ROSE} \cite{javari2020rose}), \textbf{SSNE} \cite{lu2019ssne} and methods based on matrix factorization \textbf{SPONGE} \cite{cucuringu2019sponge}. Below, we will introduce these methods separately.

\textbf{SNE} \cite{wang2017signed} is the first research on signed network embedding, which adopts the log-bilinear model (variant of Skip-Gram model) to capture both node's path and sign information. One issue present is that the relationship between the target node and multi-hop neighbor nodes is determined by the edge sign between multi-hop neighbors and the previous-hop neighbors. Specifically, for a target node $v$ and a path $h = [u_1, u_2, \cdots, u_l, v]$, the relationship between node $v$ and $u_1$ is decided by $\mathcal{E}_{u_1,u_2}$. If $\mathcal{E}_{u_1,u_2} >0$, their relationship is positive. Conversely, it is negative.

\textbf{SIGNet} \cite{islam2018signet}. This method encodes network structural information based on the word2vec model, considering both undirected and directed graph scenarios. A major innovation of this approach lies in proposing a new negative sampling method. In signed graphs, where relationships between nodes can be positive or negative, the paper suggests that nodes with positive relationships should not be included in the negative sampler set and introduces a method based on balance theory to compute the positive and negative relationships of multi-hop neighbors. In summary, neighbors connected to the target node by an even number of negative edges are assigned to the positive relationship set, while those connected by an odd number are assigned to the negative relationship set.

\textbf{SSNE} \cite{lu2019ssne} incorporates status theory into signed network embedding. In status theory, a higher status node $v_i$ has a positive link to $v_j$, while a lower status node $v_i$ has a negative link to $v_j$. SSNE represents positive links as triplets $(v_i, \mathcal{\ell}{+}, v_j)$, where the embedding space forces the combined embeddings of $h_i + h_{\mathcal{\ell}{+}}$ to be close to the embedding of $h_j$. Conversely, for negative links, the model encourages the embedding $h_i - h_{\mathcal{\ell}_{+}}$ to be close to the embedding of $h_j$.


\textbf{SIDE} \cite{kim2018side} proposed signed network embedding method builds upon truncated random walk, introducing a comprehensive likelihood formulation for signed directed connections that consistently represents both positive and negative edges. By incorporating bias factors in the likelihood function to model individual connectivity, the approach enhances the accuracy of the embedding process. Furthermore, the random walk sampling process and likelihood formulation are extended to accommodate multi-step relationships, encompassing both sign and direction. This method intricately links vector space geometry with social phenomena in networks, including homophily, preferential attachment, and balance theoretic behaviors. This association dissects the dual factors influencing link formation, establishing a robust foundation for the broad application of the method in the analysis of signed directed networks.


\textbf{ROSE} \cite{javari2020rose}. This papers attempts to address two issues. First, existing social theories fail to explain the structure in all signed graphs, meaning that not all structures conform to balance theory and status theory. Second, current signed network embedding only considers predicting the positive or negative nature of existing edges, overlooking the prediction of edge existence. To tackle these challenges, ROSE does not rely on any specific social theory for node encoding. Instead, it adopts a network transformation-based embedding approach. This involves assigning multiple roles to the same node, transforming the signed graph into an unsigned graph. Subsequently, unsigned network embedding methods can be applied to encode the graph structure.

 \textbf{SPONGE} (Signed Positive Over Negative Generalized Eigenproblem)~\cite{cucuringu2019sponge} is a $k$-way spectral clustering algorithm on signed networks. 
 Inspired by constrained clustering, a new $k$-way objective of signed graph cut is designed, which seeks the global clustering assignment for each node by minimizing the trade-off between two measures of \textit{badness}, i.e., the normalized weight of positive edges cross different clusters and the reciprocal of that of negative edges cross different clusters. 
 The new objective is cast as a regularized spectral algorithm based on solving a generalized eigenproblem. The key of solving it is to derive the goal matrix, and then perform eigenvalue decomposition on the goal matrix to obtain its eigenvectors corresponding to the $k$-smallest eigenvalues. 
 Simultaneously, it also requires that the derived embedding vectors be orthonormal for dropping the discreteness constraints. Moreover, this work also provides a detailed theoretical analysis w.r.t. the robustness of SPONGE in the Signed Stochastic Block Model framework.


\subsection{Deep Model}
Compared to shallow network embedding methods, deep models in the context of network embedding refer to those approaches that involve deep learning architectures with multiple layers in the neural network. This is currently one of the mainstream methods in SGRL.

\textbf{SiNE} \cite{wang2017signed} is the first method which employs deep learning model to learn low-dimensional vector representations for nodes of a given signed graph. 
Follow the assumption of extended structural balance theory, in embedding space, users should sit closer to their ``friends" (or users with positive links) than their ``foes" (or users with negative links). 
Based on this assumption, in order to preserve the structure information of signed graph, special triplets (e.g., $(v_i, v_j, v_k)$) (with one positive link $e_{ij} =1$ and one negative link $e_{ij} =-1$) are extracted, the learned similarity function should make sure $f(v_i, v_j) > f(v_i,v_k)$.

\textbf{SGCN} \cite{derr2018signed} tackles the challenge of applying Graph Convolutional Network (GCN) \cite{kipf2016semi} to signed networks. GCN excels in unsigned networks, but struggles with the semantic and structural complexities of negative links. SGCN introduces balance theory to categorize neighbors as positive or negative based on the number of negative edges. It maintains dual representations for each node—balanced set for positive relationships and unbalanced set for negative ones. This approach, using an extended balance theory for judging multi-hop neighbors, enables SGCN to effectively model the nuanced relationships in signed graphs.


\textbf{DNE-SBP} \cite{shen2018deep} pioneers the application of semi-supervised Stacked Autoencoder (SAE) for embedding signature networks. This deep embedding model focuses on learning low-dimensional node vector representations while preserving structural balance in signed networks. By employing semi-supervised stacked autoencoders, DNE-SBP reconstructs the adjacency connections in the given signed network. The model prioritizes sparse negative connections by applying a higher penalty to negative links during reconstruction. To maintain structural balance, pairwise constraints are incorporated, ensuring nodes connected by positive links in the embedding space are closer than those connected by negative links. This innovative approach addresses the limitation of SDNE in capturing crucial structural balance properties in signature networks.

\textbf{SNEA} \cite{li2020learning}. Attention mechanisms allow for dealing with changeable sized input and focus on the most relevant parts of input to make decisions. Graph attention network (GAT) \cite{velivckovic2017graph} is the first attention-based graph learning architecture, which assign different weights to different neighbor nodes. SNEA is a GAT-based signed network embedding methods which distinguish neighbor nodes into positive and negative neighbors which follow the similar design from \cite{derr2018signed}. The difference lies in that the contribution of neighbor nodes to the target node is not fixed; instead, by introducing attention mechanism, the model can learn an appropriate weight for each neighbor node.

\textbf{SiGAT} \cite{huang2019signed}. SiGAT is another attention-based graph neural network applied to signed networks.
Different from previously mentioned SNEA, it primarily defines different relationships by establishing various motifs. 
These motifs are related to triangles in the balance theory and status theory. 
Then, it defines an attention aggregation mechanism under different relationships, obtaining the learning of node representations in signed networks.

\textbf{SDGNN} \cite{huang2021sdgnn}. Building upon the concept of SiGAT, SDGNN further draws inspiration from GAE, treating the task of learning representations for signed graphs as an encoder-decoder framework. 
In the encoder part, SDGNN simplifies the definition of different relationships, defining only four types of signed directed relationship. 
Furthermore, it integrates the depiction of balance theory and status theory in signed networks into the  decoder parts.
SDGNN propose to recontruct the sign, direction, and trianles in loss functions.

\textbf{SGCL} \cite{shu2021sgcl} is the first to employ graph contrastive representation on signed graph. Contrastive learning is an unsupervised learning paradigm which can help model capture invariant and robust representations under perturbations. 
Compared to the unsigned graph contrastive learning framework, SGCL has three main differences. Firstly, the data augmentation method is different, taking into consideration the existence of positive and negative edges in edge relationships. Therefore, flipping positive and negative edges is used as a means of adding perturbation. Secondly, SGCL adopts the design of encoding positive and negative edges separately. Thirdly, balance theory is employed in the design of the contrastive loss function.

\textbf{SSSNET} (Semi-Supervised Signed NETwork clustering)~\cite{he2022sssnet} designed a new GNN-based aggregation mechanism, Signed Mixed-Path Aggregation (SIMPA), to aggregate up-to-$h$-hop contributing neighbors, by assigning weights for different paths. Through statistical analysis of real-world signed graph datasets\footnote{Six public datasets are used, including Sampson, Rainfall, Fine-YNet, S\&P 1500, PPI, Wiki-Rfa.}, nearly $20\%-50\%$ of triangles violate the social balance theory. Thus a variant of social balance theory is proposed to generate those paths, where a neutral stance is assumed on whether or not the enemy of an enemy is a friend. Specifically, the positive embedding is the weighted combination of the node representations from a \textit{“friend path”} where all edges need to be positive. For a target node to be an $h$-hop enemy neighbor of the source node, exactly one edge on the \textit{“enemy path”} has to be negative. Besides, a (differentiable) Probabilistic Balance Normalized Cut (PBNC)\footnote{Balance Normalized Cut (BNC) is a non-differentiable function with hard clustering assignment distribution~\cite{DBLP:conf/cikm/ChiangWD12}.} as a self-supervised loss is introduced to be optimized for training clustering on signed graphs, which minimizes the probability of negative edges assigned in intra-clusters and maximizes the probability of positive edges assigned in inter-clusters. PBNC with supervised classification loss (i.e., Cross Entropy and Triplet Loss) between the ground truths and the predicted labels is leveraged to optimize node embeddings and clustering simultaneously without any intermediate step.

\subsection{Trustworthy Model}
Trustworthy models typically refer to those representation learning methods which can produce robust, explainable ethical outcomes. Research in this area is very limited, making it a crucial direction for future studies.

\textbf{ASiNE} \cite{lee2020asine} is the first attempt that utilizes adversarial learning for signed network. Motivated by generative adversarial networks for unsigned network, ASiNE designed two pairs of generators and discriminators to generate and distinguish false positive edges and false negative edges, respectively. Two generators share an embedding space, similarly, two discriminators share another embedding space. In addition, the negative edge generator can also generate false positive edges based on the balance theory. Generators aim to generate the most indistinguishable edges, while discriminators aim to discriminate between real edges and fake edges, acting as two pairs of opponents in the following four-player minimax game. As the game progresses, the performance of both sides will gradually improve.

\textbf{RSGNN} \cite{zhang2023rsgnn} is the first paper dedicated to the robustness study of the SGNN model, investigating the impact of random noise on SGNN. The paper theoretically analyzes the limitations of the current SGNN, demonstrating that the existing SGNN model fails to learn suitable representations from unbalanced triangles. Furthermore, random noise increases the number of unbalanced triangles, thereby illustrating that the decline in SGNN performance is attributed to the introduction of random noise. Then, RSGNN explore the properties of real-world signed graph to defend the negative effect of noise and propose a novel framework RSGNN which adopts a dual architecture that simultaneously denoises the graph and learns
the node representations.

\section{Recent Advances}
Recently, the field of signed networks has seen several cutting-edge advancements, primarily including the modeling of more complex signed networks, broader applications of signed networks, and the development of tools specifically designed for signed networks.
\subsection{More Complex Signed Networks}
Compared to traditional signed social networks, researchers are increasingly focusing on more complex signed networks such as signed weighted graphs \cite{kumar2016edge}, signed temporal graphs \cite{sharma2023representation}, signed hypergraphs \cite{chen2020neural} and so on. 
Within these domains, the fields of signed bipartite graphs modeling object opinions and signed temporal graphs modeling temporal dynamics have seen the emergence of some typical works, such as \textbf{SBGNN}, \textbf{SBGCL}, and \textbf{SEMBA}.

\textbf{SBGNN} \cite{huang2021signed} is a SGNN model designed for signed bipartite graphs. Like most GNNs message-passing scheme, SBGNNs follow a message-passing scheme.
However, in SBGNN, new message functions, aggregation functions, and update functions are defined by applying balance theory. More specifically, message neighborhood propagation is divided into four categories. 
SBGNN captures the higher-order information through a layer-by-layer design.

\textbf{SBGCL} \cite{zhang2023contrastive} is another model that applies contrastive learning models to signed bipartite graph representation learning. 
Unlike SGCL, SBGCL focuses on signed bipartite graphs, attempting to address the issue of SGCL's inability to capture potential relationships among nodes of the same type. 
SBGCL enhances a signed bipartite graph through an innovative two-level graph augmentation method and a multi-perspective contrastive loss is employed to unify the node presentations learned from the two perspectives. 

\textbf{SEMBA} \cite{sharma2023representation} is a signed GNN model that combine sign and dynamci in social networks.  In order to addressing temporal-awareness, staleness, and sign-awareness problems, SEMBA use  memory modules and balanced aggregation to learn short-term memory encoding and a long-term embedding. More specifically, memories contains both positive and negative parts, and aggregation encode positive and negative embeddings following balance theory.

\subsection{Applications}
With the rise of social media, social conflicts have intensified, manifesting in an increase in negative links between individuals. 
In this section, we provide an overview of some of the most attention-grabbing applications currently.

\textbf{Social Polarization} 
Using signed graphs to study adverse effects is a promising research topic. These adverse effects, such as polarization and echo chambers, can be harmful to the process of democratic deliberation in our society\cite{xiao2020searching,bonchi2019discovering}. 
For example, \textbf{POLE} \cite{huang2022pole} is a signed networking embedding methods, adopting a measure of polarization based on the signed random-walks. 
\textbf{POLE} first design a polarization measure for signed graphs and use matrix factorization optimized polarized similarity consistency.

\textbf{Stance Detection}
Stance detection typically categorizes stances into three types: ``support," ``oppose," and ``neutral," where in this context, ``support" can be viewed as a positive relationship and ``oppose" as a negative relationship\footnote{Neutral could be modeled as either a positive relationship or unknown.}. 
While a substantial amount of work on stance detection is based on Natural Language Processing (NLP). 
However, an increasing number of studies have found that social networks significantly influence people's opinions on specific matters \cite{pougue2023learning}.
SEM \cite{pougue2023learning} jointly learns user and topic embeddings in signed social graphs with distinct edge types for each topic.

\textbf{Recommendation System with Negative Feedback}
With the success of graph neural network models in the field of recommendation systems, these systems can be viewed as user-item interaction bipartite graphs.
A substantial number of researchers have employed graph representation learning to model these bipartite graphs, as seen in works like PinSAGE \cite{ying2018graph} and LightGCN \cite{he2020lightgcn}.
However, in the context of recommendation systems, the negative feedback mechanism \cite{xie2021deep}  also refers to users expressing dissatisfaction with recommended products, manifested through actions like poor reviews, skipping recommended content, or reporting. 
The optimization of recommendation systems using negative feedback has garnered attention from researchers \cite{tang2016recommendations}. 
Notably, \cite{seo2022siren} introduced the \textbf{SiReN} model, which employs a signed graph neural network approach to model recommendation systems incorporating both positive and negative feedback. 
By conducting qualitative and quantitative experiments on the mechanism of negative feedback, \cite{huang2023negative} found that negative feedback interactions (i.e., negative edges) could enhance the performance of recommendation systems (specifically, the prediction of positive edges) to a certain extent. 
They proposed a new Signed GNN model named \textbf{SiGRec} with a new SiC loss function to model signed graphs in recommendation systems.

\textbf{Social Computing} 
In addition to the applications mentioned above, due to the relevance of its links to human sentiment, SGRL has also been applied in various social computing  fields. 
These include education \cite{ni2023enhancing}, communication \cite{he2022positive}, cryptocurrency, politics, and more. 
This broad applicability underscores the significance of understanding and leveraging the dynamics of signed networks in diverse social scenarios.

\subsection{Tools}
With the rise of graph representation learning, an increasing number of tools for modeling graph representations have been introduced. 
Among the most representative of these are Pytorch Geometric (PyG) and Deep Graph Library (DGL).
However, these approaches often view signed graph merely as a type of relation, overlooking the social attributes and real-world contexts of signed networks. 
PyTorch Geometric Signed Directed (\textbf{PyGSD}) \cite{he2022pytorch} takes data classes, data loaders, as and data splitters into consideration, and design  a specific tool for SGRL.

\section{Future Direction}

SGRL is a significant branch of graph representation learning. Despite notable advancements in this field, there still exists a plethora of research directions worthy of exploration. 

\textbf{Trustworthy research.} GNNs focused on performance have shown potential drawbacks such as susceptibility to adversarial attacks, unexplainable bias against disadvantaged groups, and excessive resource consumption in edge computing settings. To mitigate these unintended consequences, the imperative is to construct proficient GNNs distinguished by their trustworthiness. For signed graph representation learning, current related research is limited to adversarial attack and defense \cite{zhou2023black,zhang2023rsgnn}. There is currently no relevant research in other crucial areas of trustworthy Graph Neural Networks (GNNs), such as interpretability, privacy, fairness, and accountability. Taking explainability research as an example, explainability research is to enhance understanding and trust in complex deep models by providing interpretable insights into their decision-making processes and predictions. Several explainability methods have been introduced to elucidate the underlying mechanisms of unsigned graph neural networks \cite{yuan2022explainability}. However, there is currently no research on the interpretability of the SGNN model. Compared to explainability research on Graph Neural Networks (GNN), Signed GNN possesses two new features, new downstream task (i.e., link sign prediction), and new collective properties (i.e., balance theory and status theory). Therefore, considering the combination of the above two types of features, we believe that designing explainability methods for SGNN is a promising research direction.



\textbf{Data-centric research.} Data-centric AI is the discipline of systematically engineering the data used to build an AI system (by Andrew Ng), which has a broad research content, e.g., data collection, data labeling, data preparation, data augmentation, etc. Graph data augmentation \cite{liu2022local} is undoubtedly the most popular direction in the field currently. These graph data augmentation methods often rely on side information such as node labels and features, making it challenging to directly apply them to enhance signed graph data. However, for signed graphs, the current research efforts are still quite limited \cite{zhang2023sga}. Sparse labels represent another prominent challenge in signed graph representation learning. This challenge is also expected to be addressed in data-centric research.



\textbf{Large Language Models (LLMs).} Large language models are currently a hot topic in research. In contrast to non-contextualized shallow textual embeddings, large language models (LLMs) exhibit extensive context-aware knowledge and superior semantic comprehension capabilities, achieved through pre-training on vast text corpora. The utilization of Large Language Models (LLMs) in Graph Representation Learning has seen some attempts in the past year \cite{chen2023exploring}. However, without a doubt, research in this domain is still at a very early stage. The exploration of how to apply LLMs to signed graph representation learning is an intriguing area, and as of now, there is no existing research in this specific domain.

\clearpage
\bibliographystyle{named}
\bibliography{refs}

\begin{thebibliography}{}

\bibitem[\protect\citeauthoryear{Bonchi \bgroup \em et al.\egroup
  }{2019}]{bonchi2019discovering}
Francesco Bonchi, Edoardo Galimberti, Aristides Gionis, Bruno Ordozgoiti, and
  Giancarlo Ruffo.
\newblock Discovering polarized communities in signed networks.
\newblock In {\em CIKM}, 2019.

\bibitem[\protect\citeauthoryear{Cai \bgroup \em et al.\egroup
  }{2018}]{cai2018comprehensive}
Hongyun Cai, Vincent~W Zheng, and Kevin Chen-Chuan Chang.
\newblock A comprehensive survey of graph embedding: Problems, techniques, and
  applications.
\newblock {\em TKDE}, 2018.

\bibitem[\protect\citeauthoryear{Chakraborty \bgroup \em et al.\egroup
  }{2023}]{chakraborty2023siggan}
Roshni Chakraborty, Ritwika Das, and Joydeep Chandra.
\newblock Siggan: Adversarial model for learning signed relationships in
  networks.
\newblock {\em TKDD}, 2023.

\bibitem[\protect\citeauthoryear{Chen \bgroup \em et al.\egroup
  }{2018}]{chen2018bridge}
Yiqi Chen, Tieyun Qian, Huan Liu, and Ke~Sun.
\newblock "bridge": Enhanced signed directed network embedding.
\newblock In {\em CIKM}, 2018.

\bibitem[\protect\citeauthoryear{Chen \bgroup \em et al.\egroup
  }{2020}]{chen2020neural}
Xu~Chen, Kun Xiong, Yongfeng Zhang, Long Xia, Dawei Yin, and Jimmy~Xiangji
  Huang.
\newblock Neural feature-aware recommendation with signed hypergraph
  convolutional network.
\newblock {\em TOIS}, 2020.

\bibitem[\protect\citeauthoryear{Chen \bgroup \em et al.\egroup
  }{2023}]{chen2023exploring}
Zhikai Chen, Haitao Mao, Hang Li, Wei Jin, Hongzhi Wen, Xiaochi Wei, Shuaiqiang
  Wang, Dawei Yin, Wenqi Fan, Hui Liu, et~al.
\newblock Exploring the potential of large language models (llms) in learning
  on graphs.
\newblock {\em arXiv preprint arXiv:2307.03393}, 2023.

\bibitem[\protect\citeauthoryear{Chiang \bgroup \em et al.\egroup
  }{2012}]{DBLP:conf/cikm/ChiangWD12}
Kai{-}Yang Chiang, Joyce~Jiyoung Whang, and Inderjit~S. Dhillon.
\newblock Scalable clustering of signed networks using balance normalized cut.
\newblock In {\em CIKM}, 2012.

\bibitem[\protect\citeauthoryear{Cucuringu \bgroup \em et al.\egroup
  }{2019}]{cucuringu2019sponge}
Mihai Cucuringu, Peter Davies, Aldo Glielmo, and Hemant Tyagi.
\newblock {SPONGE:} {A} generalized eigenproblem for clustering signed
  networks.
\newblock In {\em AISTATS}, 2019.

\bibitem[\protect\citeauthoryear{Cui \bgroup \em et al.\egroup
  }{2018}]{cui2018survey}
Peng Cui, Xiao Wang, Jian Pei, and Wenwu Zhu.
\newblock A survey on network embedding.
\newblock {\em TKDE}, 2018.

\bibitem[\protect\citeauthoryear{Derr \bgroup \em et al.\egroup
  }{2018}]{derr2018signed}
Tyler Derr, Yao Ma, and Jiliang Tang.
\newblock Signed graph convolutional networks.
\newblock In {\em ICDM}, 2018.

\bibitem[\protect\citeauthoryear{Hamilton \bgroup \em et al.\egroup
  }{2017}]{hamilton2017inductive}
William~L. Hamilton, Zhitao Ying, and Jure Leskovec.
\newblock Inductive representation learning on large graphs.
\newblock In {\em NeurIPS}, 2017.

\bibitem[\protect\citeauthoryear{He \bgroup \em et al.\egroup
  }{2020}]{he2020lightgcn}
Xiangnan He, Kuan Deng, Xiang Wang, Yan Li, Yong{-}Dong Zhang, and Meng Wang.
\newblock Lightgcn: Simplifying and powering graph convolution network for
  recommendation.
\newblock In {\em SIGIR}, 2020.

\bibitem[\protect\citeauthoryear{He \bgroup \em et al.\egroup
  }{2022a}]{he2022positive}
Qiang He, Hongwei Du, and Ziwei Liang.
\newblock Positive influence maximization in signed networks within a limited
  time.
\newblock {\em IEEE TCSS}, 2022.

\bibitem[\protect\citeauthoryear{He \bgroup \em et al.\egroup
  }{2022b}]{he2022sssnet}
Yixuan He, Gesine Reinert, Songchao Wang, and Mihai Cucuringu.
\newblock Sssnet: Semi-supervised signed network clustering.
\newblock In {\em SDM}, 2022.

\bibitem[\protect\citeauthoryear{He \bgroup \em et al.\egroup
  }{2023}]{he2022pytorch}
Yixuan He, Xitong Zhang, Junjie Huang, Benedek Rozemberczki, Mihai Cucuringu,
  and Gesine Reinert.
\newblock {PyTorch Geometric Signed Directed: A Software Package on Graph
  Neural Networks for Signed and Directed Graphs}.
\newblock In {\em LoG}, 2023.

\bibitem[\protect\citeauthoryear{Heider}{1946}]{heider1946attitudes}
Fritz Heider.
\newblock Attitudes and cognitive organization.
\newblock {\em The Journal of psychology}, 1946.

\bibitem[\protect\citeauthoryear{Huang \bgroup \em et al.\egroup
  }{2019}]{huang2019signed}
Junjie Huang, Huawei Shen, Liang Hou, and Xueqi Cheng.
\newblock Signed graph attention networks.
\newblock In {\em ICANN}, 2019.

\bibitem[\protect\citeauthoryear{Huang \bgroup \em et al.\egroup
  }{2021a}]{huang2021signed}
Junjie Huang, Huawei Shen, Qi~Cao, Shuchang Tao, and Xueqi Cheng.
\newblock Signed bipartite graph neural networks.
\newblock In {\em CIKM}, 2021.

\bibitem[\protect\citeauthoryear{Huang \bgroup \em et al.\egroup
  }{2021b}]{huang2021sdgnn}
Junjie Huang, Huawei Shen, Liang Hou, and Xueqi Cheng.
\newblock {SDGNN:} learning node representation for signed directed networks.
\newblock In {\em AAAI}, 2021.

\bibitem[\protect\citeauthoryear{Huang \bgroup \em et al.\egroup
  }{2022}]{huang2022pole}
Zexi Huang, Arlei Silva, and Ambuj Singh.
\newblock Pole: Polarized embedding for signed networks.
\newblock In {\em WSDM}, 2022.

\bibitem[\protect\citeauthoryear{Huang \bgroup \em et al.\egroup
  }{2023}]{huang2023negative}
Junjie Huang, Ruobing Xie, Qi~Cao, Huawei Shen, Shaoliang Zhang, Feng Xia, and
  Xueqi Cheng.
\newblock Negative can be positive: Signed graph neural networks for
  recommendation.
\newblock {\em IPM}, 2023.

\bibitem[\protect\citeauthoryear{Islam \bgroup \em et al.\egroup
  }{2018}]{islam2018signet}
Mohammad~Raihanul Islam, B~Aditya~Prakash, and Naren Ramakrishnan.
\newblock Signet: Scalable embeddings for signed networks.
\newblock In {\em PAKDD}. Springer, 2018.

\bibitem[\protect\citeauthoryear{Javari \bgroup \em et al.\egroup
  }{2020}]{javari2020rose}
Amin Javari, Tyler Derr, Pouya Esmailian, Jiliang Tang, and Kevin~Chen{-}Chuan
  Chang.
\newblock {ROSE:} role-based signed network embedding.
\newblock In {\em WWW}, 2020.

\bibitem[\protect\citeauthoryear{Kim \bgroup \em et al.\egroup
  }{2018}]{kim2018side}
Junghwan Kim, Haekyu Park, Ji{-}Eun Lee, and U~Kang.
\newblock {SIDE:} representation learning in signed directed networks.
\newblock In {\em WWW}, 2018.

\bibitem[\protect\citeauthoryear{Kipf and Welling}{2017}]{kipf2016semi}
Thomas~N. Kipf and Max Welling.
\newblock Semi-supervised classification with graph convolutional networks.
\newblock In {\em ICLR}, 2017.

\bibitem[\protect\citeauthoryear{Kumar \bgroup \em et al.\egroup
  }{2016}]{kumar2016edge}
Srijan Kumar, Francesca Spezzano, VS~Subrahmanian, and Christos Faloutsos.
\newblock Edge weight prediction in weighted signed networks.
\newblock In {\em ICDM}, 2016.

\bibitem[\protect\citeauthoryear{Lamb \bgroup \em et al.\egroup
  }{2020}]{lamb2020graph}
Lu{\'{\i}}s~C. Lamb, Artur~S. d'Avila Garcez, Marco Gori, Marcelo O.~R. Prates,
  Pedro H.~C. Avelar, and Moshe~Y. Vardi.
\newblock Graph neural networks meet neural-symbolic computing: {A} survey and
  perspective.
\newblock In {\em IJCAI}, 2020.

\bibitem[\protect\citeauthoryear{Lee \bgroup \em et al.\egroup
  }{2020}]{lee2020asine}
Yeon{-}Chang Lee, Nayoun Seo, Kyungsik Han, and Sang{-}Wook Kim.
\newblock {ASiNE}: Adversarial signed network embedding.
\newblock In {\em SIGIR}, 2020.

\bibitem[\protect\citeauthoryear{Leskovec \bgroup \em et al.\egroup
  }{2010}]{leskovec2010signed}
Jure Leskovec, Daniel~P. Huttenlocher, and Jon~M. Kleinberg.
\newblock Signed networks in social media.
\newblock In {\em CHI}, 2010.

\bibitem[\protect\citeauthoryear{Li \bgroup \em et al.\egroup
  }{2019}]{li2019senti2pop}
Jinning Li, Yirui Gao, Xiaofeng Gao, Yan Shi, and Guihai Chen.
\newblock Senti2pop: sentiment-aware topic popularity prediction on social
  media.
\newblock In {\em ICDM}, 2019.

\bibitem[\protect\citeauthoryear{Li \bgroup \em et al.\egroup
  }{2020}]{li2020learning}
Yu~Li, Yuan Tian, Jiawei Zhang, and Yi~Chang.
\newblock Learning signed network embedding via graph attention.
\newblock In {\em AAAI}, 2020.

\bibitem[\protect\citeauthoryear{Liu \bgroup \em et al.\egroup
  }{2021}]{liu2021signed}
Haoxin Liu, Ziwei Zhang, Peng Cui, Yafeng Zhang, Qiang Cui, Jiashuo Liu, and
  Wenwu Zhu.
\newblock Signed graph neural network with latent groups.
\newblock In {\em KDD}, 2021.

\bibitem[\protect\citeauthoryear{Liu \bgroup \em et al.\egroup
  }{2022}]{liu2022local}
Songtao Liu, Rex Ying, Hanze Dong, Lanqing Li, Tingyang Xu, Yu~Rong, Peilin
  Zhao, Junzhou Huang, and Dinghao Wu.
\newblock Local augmentation for graph neural networks.
\newblock In {\em ICML}, 2022.

\bibitem[\protect\citeauthoryear{Liu \bgroup \em et al.\egroup
  }{2023}]{liu2022graph}
Chuang Liu, Yibing Zhan, Jia Wu, Chang Li, Bo~Du, Wenbin Hu, Tongliang Liu, and
  Dacheng Tao.
\newblock Graph pooling for graph neural networks: Progress, challenges, and
  opportunities.
\newblock In {\em IJCAI}, 2023.

\bibitem[\protect\citeauthoryear{Lu \bgroup \em et al.\egroup
  }{2019}]{lu2019ssne}
Chunyu Lu, Pengfei Jiao, Hongtao Liu, Yaping Wang, Hongyan Xu, and Wenjun Wang.
\newblock Ssne: status signed network embedding.
\newblock In {\em PAKDD}, 2019.

\bibitem[\protect\citeauthoryear{Ni \bgroup \em et al.\egroup
  }{2023}]{ni2023enhancing}
Lin Ni, Sijie Wang, Zeyu Zhang, Xiaoxuan Li, Xianda Zheng, Paul Denny, and
  Jiamou Liu.
\newblock Enhancing student performance prediction on learnersourced questions
  with sgnn-llm synergy.
\newblock {\em ArXiv preprint}, 2023.

\bibitem[\protect\citeauthoryear{Pougu{\'e}-Biyong \bgroup \em et al.\egroup
  }{2023}]{pougue2023learning}
John Pougu{\'e}-Biyong, Akshay Gupta, Aria Haghighi, and Ahmed El-Kishky.
\newblock {Learning stance embeddings from signed social graphs}.
\newblock In {\em WSDM}, 2023.

\bibitem[\protect\citeauthoryear{Seo \bgroup \em et al.\egroup
  }{2022}]{seo2022siren}
Changwon Seo, Kyeong-Joong Jeong, Sungsu Lim, and Won-Yong Shin.
\newblock Siren: Sign-aware recommendation using graph neural networks.
\newblock {\em TNNLS}, 2022.

\bibitem[\protect\citeauthoryear{Sharma \bgroup \em et al.\egroup
  }{2023}]{sharma2023representation}
Kartik Sharma, Mohit Raghavendra, Yeon-Chang Lee, and Srijan Kumar.
\newblock Representation learning in continuous-time dynamic signed networks.
\newblock In {\em CIKM}, 2023.

\bibitem[\protect\citeauthoryear{Shen and Chung}{2018}]{shen2018deep}
Xiao Shen and Fu-Lai Chung.
\newblock Deep network embedding for graph representation learning in signed
  networks.
\newblock {\em IEEE Transactions on Cybernetics}, 2018.

\bibitem[\protect\citeauthoryear{Shu \bgroup \em et al.\egroup
  }{2021}]{shu2021sgcl}
Lin Shu, Erxin Du, Yaomin Chang, Chuan Chen, Zibin Zheng, Xingxing Xing, and
  Shaofeng Shen.
\newblock Sgcl: Contrastive representation learning for signed graphs.
\newblock In {\em CIKM}, 2021.

\bibitem[\protect\citeauthoryear{Sun \bgroup \em et al.\egroup
  }{2020}]{sun2020stable}
Renjie Sun, Chen Chen, Xiaoyang Wang, Ying Zhang, and Xun Wang.
\newblock Stable community detection in signed social networks.
\newblock {\em TKDE}, (10), 2020.

\bibitem[\protect\citeauthoryear{Tang \bgroup \em et al.\egroup
  }{2016a}]{tang2016recommendations}
Jiliang Tang, Charu~C. Aggarwal, and Huan Liu.
\newblock Recommendations in signed social networks.
\newblock In {\em WWW}, 2016.

\bibitem[\protect\citeauthoryear{Tang \bgroup \em et al.\egroup
  }{2016b}]{tang2016survey}
Jiliang Tang, Yi~Chang, Charu Aggarwal, and Huan Liu.
\newblock A survey of signed network mining in social media.
\newblock {\em ACM CSUR}, 2016.

\bibitem[\protect\citeauthoryear{Velickovic \bgroup \em et al.\egroup
  }{2018}]{velivckovic2017graph}
Petar Velickovic, Guillem Cucurull, Arantxa Casanova, Adriana Romero, Pietro
  Li{\`{o}}, and Yoshua Bengio.
\newblock Graph attention networks.
\newblock In {\em ICLR}, 2018.

\bibitem[\protect\citeauthoryear{Wang \bgroup \em et al.\egroup
  }{2017}]{wang2017signed}
Suhang Wang, Jiliang Tang, Charu~C. Aggarwal, Yi~Chang, and Huan Liu.
\newblock Signed network embedding in social media.
\newblock In {\em SDM}, 2017.

\bibitem[\protect\citeauthoryear{Wang \bgroup \em et al.\egroup
  }{2018}]{wang2018shine}
Hongwei Wang, Fuzheng Zhang, Min Hou, Xing Xie, Minyi Guo, and Qi~Liu.
\newblock {SHINE:} signed heterogeneous information network embedding for
  sentiment link prediction.
\newblock In {\em WSDM}, 2018.

\bibitem[\protect\citeauthoryear{Wu \bgroup \em et al.\egroup
  }{2020a}]{wu2020graph}
Shiwen Wu, Fei Sun, Wentao Zhang, Xu~Xie, and Bin Cui.
\newblock Graph neural networks in recommender systems: a survey.
\newblock {\em ACM CSUR}, 2020.

\bibitem[\protect\citeauthoryear{Wu \bgroup \em et al.\egroup
  }{2020b}]{wu2020comprehensive}
Zonghan Wu, Shirui Pan, Fengwen Chen, Guodong Long, Chengqi Zhang, and S~Yu
  Philip.
\newblock A comprehensive survey on graph neural networks.
\newblock {\em TNNLS}, 2020.

\bibitem[\protect\citeauthoryear{Xiao \bgroup \em et al.\egroup
  }{2020}]{xiao2020searching}
Han Xiao, Bruno Ordozgoiti, and Aristides Gionis.
\newblock Searching for polarization in signed graphs: a local spectral
  approach.
\newblock In {\em WWW}, 2020.

\bibitem[\protect\citeauthoryear{Xie \bgroup \em et al.\egroup
  }{2020}]{xie2021deep}
Ruobing Xie, Cheng Ling, Yalong Wang, Rui Wang, Feng Xia, and Leyu Lin.
\newblock Deep feedback network for recommendation.
\newblock In {\em IJCAI}, 2020.

\bibitem[\protect\citeauthoryear{Ying \bgroup \em et al.\egroup
  }{2018}]{ying2018graph}
Rex Ying, Ruining He, Kaifeng Chen, Pong Eksombatchai, William~L. Hamilton, and
  Jure Leskovec.
\newblock Graph convolutional neural networks for web-scale recommender
  systems.
\newblock In {\em KDD}, 2018.

\bibitem[\protect\citeauthoryear{Yuan \bgroup \em et al.\egroup
  }{2017}]{yuan2017sne}
Shuhan Yuan, Xintao Wu, and Yang Xiang.
\newblock {SNE: Signed Network Embedding}.
\newblock In {\em PAKDD}, 2017.

\bibitem[\protect\citeauthoryear{Yuan \bgroup \em et al.\egroup
  }{2022}]{yuan2022explainability}
Hao Yuan, Haiyang Yu, Shurui Gui, and Shuiwang Ji.
\newblock Explainability in graph neural networks: A taxonomic survey.
\newblock {\em TPAMI}, 2022.

\bibitem[\protect\citeauthoryear{Zhang \bgroup \em et al.\egroup
  }{2020}]{zhang2020deep}
Ziwei Zhang, Peng Cui, and Wenwu Zhu.
\newblock Deep learning on graphs: A survey.
\newblock {\em TKDE}, 2020.

\bibitem[\protect\citeauthoryear{Zhang \bgroup \em et al.\egroup
  }{2023a}]{zhang2023contrastive}
Zeyu Zhang, Jiamou Liu, Kaiqi Zhao, Song Yang, Xianda Zheng, and Yifei Wang.
\newblock Contrastive learning for signed bipartite graphs.
\newblock In {\em SIGIR}, 2023.

\bibitem[\protect\citeauthoryear{Zhang \bgroup \em et al.\egroup
  }{2023b}]{zhang2023rsgnn}
Zeyu Zhang, Jiamou Liu, Xianda Zheng, Yifei Wang, Pengqian Han, Yupan Wang,
  Kaiqi Zhao, and Zijian Zhang.
\newblock Rsgnn: A model-agnostic approach for enhancing the robustness of
  signed graph neural networks.
\newblock In {\em Web Conference}, 2023.

\bibitem[\protect\citeauthoryear{Zhang \bgroup \em et al.\egroup
  }{2023c}]{zhang2023sga}
Zeyu Zhang, Shuyan Wan, Sijie Wang, Xianda Zheng, Xinrui Zhang, Kaiqi Zhao,
  Jiamou Liu, and Dong Hao.
\newblock Sga: A graph augmentation method for signed graph neural networks.
\newblock {\em ArXiv preprint}, 2023.

\bibitem[\protect\citeauthoryear{Zheng \bgroup \em et al.\egroup
  }{2015}]{zheng2015social}
Xiaolong Zheng, Daniel Zeng, and Fei-Yue Wang.
\newblock Social balance in signed networks.
\newblock {\em Information Systems Frontiers}, 2015.

\bibitem[\protect\citeauthoryear{Zhou \bgroup \em et al.\egroup
  }{2020}]{zhou2020graph}
Jie Zhou, Ganqu Cui, Shengding Hu, Zhengyan Zhang, Cheng Yang, Zhiyuan Liu,
  Lifeng Wang, Changcheng Li, and Maosong Sun.
\newblock Graph neural networks: A review of methods and applications.
\newblock {\em AI Open}, 2020.

\bibitem[\protect\citeauthoryear{Zhou \bgroup \em et al.\egroup
  }{2023}]{zhou2023black}
Jialong Zhou, Yuni Lai, Jian Ren, and Kai Zhou.
\newblock Black-box attacks against signed graph analysis via balance
  poisoning.
\newblock {\em ArXiv preprint}, 2023.

\bibitem[\protect\citeauthoryear{Zitnik \bgroup \em et al.\egroup
  }{2018}]{zitnik2018modeling}
Marinka Zitnik, Monica Agrawal, and Jure Leskovec.
\newblock Modeling polypharmacy side effects with graph convolutional networks.
\newblock {\em Bioinformatics}, 2018.

\end{thebibliography}

\end{document}